\documentclass[prb,twocolumn,showpacs]{revtex4}
\usepackage{graphicx} 

\usepackage{amsmath,amsthm,amssymb,mathrsfs}
\usepackage{bm}

\begin{document}

\title{Phase estimation of spin-torque oscillator by nonlinear spin-torque diode effect}

\author{Terufumi Yamaguchi, Sumito Tsunegi, and Tomohiro Taniguchi}
\affiliation{National Institute of Advanced Industrial Science and Technology (AIST), Spintronics Research Center, Tsukuba, Ibaraki 305-8568, Japan}

\date{\today} 
\begin{abstract}
  {
A theoretical analysis is developed on spin-torque diode effect in nonlinear region. 
An analytical solution of the diode voltage generated from spin-torque oscillator by the rectification of an alternating current is derived. 
The diode voltage is revealed to depend nonlinearly on the phase difference between the oscillator and the alternating current. 
The validity of the analytical prediction is confirmed by numerical simulation of the Landau-Lifshitz-Gilbert equation. 
The results indicate that the spin-torque diode effect is useful to evaluate the phase of a spin-torque oscillator in forced synchronization state. 
  }
\end{abstract}

\maketitle


Generating microwave power by using spin-torque oscillator (STO) \cite{kiselev03,rippard04,krivorotov05,houssameddine07,urazhdin10,kubota13} has been an exciting topic in the field of spintronics 
because of the applicability to practical devices such as magnetic recording head \cite{zhu08,kudo10,bosu16,suto17}. 
The previous works on STO have focused on its frequency, linewidth, and/or power 
because these quantities determine the quality of the STO assembled in microwave generators. 
Recent growth of interest on the applicability of STOs to other technologies, 
such as neuromorphic computing and phased array radar \cite{torrejon17,kudo17,furuta18,tsunegi18SR,tsunegi18JJAP,markovic19,tsunegi19}, 
motivates us to investigate another physical quantity of the oscillator, namely phase. 
For example, the pattern recognition by using an array of spin-Hall oscillators is based on the control of the phase differences among the oscillators \cite{kudo17}. 
The performance of the reservoir computing was improved by using the phase synchronization of an STO to a microwave magnetic field \cite{markovic19,tsunegi19}. 
The phased array radar controls the propagating direction of the wave signal by tuning the phase difference between the oscillators and the signal. 
As can be seen in these examples, the phase plays a key role in next-generation spintronics devices. 
However, studies investigating the STO's phase are still few \cite{rippard05,zhou08,finocchio12}. 
In this work, we focus on the phase difference of an STO in an injection-locked (forced synchronization) state, 
where the oscillation frequency and phase of the STO are locked to those of an injected alternating current. 


Spin-torque diode \cite{tulapurkar05,kubota08,sankey08} is another spintronics device generating a direct voltage by rectifying an injected alternating current. 
The spin-torque diode effect is caused by a linear (small amplitude) oscillation of the magnetization. 
Recently, however, the spin-torque diode effect has been extended to nonlinear region \cite{cheng13,fang16,zhang18,comment_suzuki}. 
It should be emphasized here that the spin-torque diode effect in the nonlinear region corresponds to the injection locking of an STO; see also the description below. 
Although the previous works partially implied that the diode voltage in the nonlinear region reflects the phase of the STO, 
the main focus was on the experiments to enhance the diode sensitivity. 
A detail analysis of the relation between the diode voltage and the phase of the STO has not been developed yet from the theoretical point of view.

In this work, we have developed a theoretical framework proposing an evaluation method of the STO's phase 
in an injection-locked state by focusing on the spin-torque diode signal from the STO. 
It is analytically shown that the spin-torque diode voltage of the STO in the frequency domain 
depends nonlinearly on the phase difference between the oscillator and injected alternating current. 
Numerical simulation of the Landau-Lifshitz-Gilbert (LLG) equation is also performed to confirm the analytical prediction. 
The results indicate that the spin-torque diode measurement in nonlinear region can be used as a convenient experimental tool to evaluate the phase of the STO. 


Before showing our calculation details, let us first emphasize the difference of the spin-torque diode effect between the linear and nonlinear regions. 
The conventional spin-torque diode effect \cite{tulapurkar05,kubota08,sankey08} is a linear effect. 
It is caused by an alternating current, and is related to a linear oscillation of the magnetization called ferromagnetic resonance (FMR). 
The output is the direct voltage as a result of the rectification of the alternating current, 
and has a peak at the FMR frequency. 
When a direct current is simultaneously injected into the diode, it results in a modulation of the spectrum linewidth. 
Note however that the necessity of the direct current is, in principle, not essential in the linear spin-torque diode effect. 

On the other hand, the spin-torque diode effect in the nonlinear region in this work corresponds to the injection locking of an STO. 
The auto-oscillation in the STO is a nonlinear oscillation caused by an injection of a direct current. 
The oscillation frequency of the STO can be tuned by changing the magnitude of the direct current. 
The output of the STO is presented as an oscillating power. 
When an alternating current is simultaneously injected into the STO with some conditions fulfilled, however, 
the oscillation frequency and phase of the STO are locked to those of the alternating current. 
The phenomenon is called the injection locking or forced synchronization. 
Note that, because of the presence of the alternating current, 
the STO in the injection-locking state is also expected to output a direct (rectified) voltage, 
similar to the conventional spin-torque diode effect. 
The direct voltage is calculated in the following. 




\begin{figure}
\centerline{\includegraphics[width=1.0\columnwidth]{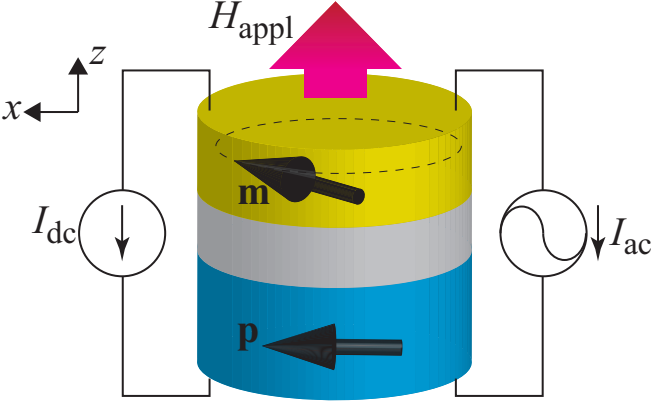}}
\caption{
        Schematic view of the spin torque oscillator with the perpendicularly magnetized free layer and in-plane magnetized reference layer. 
        The unit vectors $\mathbf{m}$ and $\mathbf{p}$ point to the directions of the magnetizations of the free and reference layers. 
        Direct and alternating currents are applied to the oscillator, in addition to a perpendicular magnetic field $H_{\rm appl}$ in the $z$ direction. 
         \vspace{-3ex}}
\label{fig:fig1}
\end{figure}



We consider an STO consisting of a perpendicularly magnetized free layer and an in-plane magnetized reference layer \cite{kubota13} schematically shown in Fig. \ref{fig:fig1}. 
The $z$ axis is perpendicular to the film-plane, 
whereas the $x$ axis is parallel to the magnetization direction of the reference layer. 
An external field $H_{\rm appl}$ and electric current $I$ is applied along the $z$ direction. 
A positive current corresponds to the electrons flowing from the free to reference layer. 
In the present work, the current consists of direct and alternating currents as 
\begin{equation}
  I
  =
  I_{\rm dc}
  +
  I_{\rm ac}
  \cos 2\pi f_{\rm ac} t, 
  \label{eq:current_def}
\end{equation}
where $I_{\rm dc}$ and $I_{\rm ac}$ are the amplitudes of the direct and alternating currents, 
whereas $f_{\rm ac}$ corresponds to the frequency of the alternating current. 
It was clarified in the previous works that the magnetization dynamics in the STO is well described by the LLG equation with macrospin assumption \cite{kubota13,taniguchi13APEX}, 
which is given by 
\begin{equation}
  \frac{d \mathbf{m}}{dt}
  =
  -\gamma
  \mathbf{m}
  \times
  \mathbf{H}
  -
  \gamma
  H_{\rm s}
  \mathbf{m}
  \times
  \left(
    \mathbf{p}
    \times
    \mathbf{m}
  \right)
  +
  \alpha 
  \mathbf{m}
  \times
  \frac{d \mathbf{m}}{dt},
  \label{eq:LLG}
\end{equation}
where $\mathbf{m}$ and $\mathbf{p}$ are the unit vectors pointing in the magnetization directions of the free and reference layers, respectively. 
The gyromagnetic ratio and the Gilbert damping constant are denoted as $\gamma$ and $\alpha$, respectively. 
The magnetic field $\mathbf{H}$ consists of the perpendicular field $H_{\rm appl}$, the interfacial anisotropy field $H_{\rm K}$, and the demagnetization field $-4\pi M$ as 
\begin{equation}
  \mathbf{H}
  =
  \left[
    H_{\rm appl}
    +
    \left(
      H_{\rm K}
      -
      4\pi M
    \right)
    m_{z}
  \right]
  \mathbf{e}_{z}. 
  \label{eq:field}
\end{equation}
The spin-torque strength is given by 
\begin{equation}
  H_{\rm s}
  =
  \frac{\hbar \eta I}{2e(1+\lambda \mathbf{m}\cdot\mathbf{p})MV}, 
\end{equation}
where $\eta$ is the spin polarization of the current whereas $\lambda$ characterizes the angular dependence of the spin torque \cite{slonczewski05}. 
The saturation magnetization and volume of the free layer are denoted as $M$ and $V$, respectively. 
It is useful to introduce 
$H_{\rm ac}=\hbar \eta I_{\rm ac}/(2eMV)$ for the latter discussion, 
which represents the magnitude of the contribution from the alternating current to the spin torque. 




Let us first consider the auto-oscillation in the absence of the alternating current. 
We introduce the zenith and azimuth angles, $(\theta,\varphi)$, as $\mathbf{m}=(\sin\theta\cos\varphi,\sin\theta\sin\varphi,\cos\theta)$. 
In the auto-oscillation state, the angle $\theta$ is almost constant, as clarified in our previous work \cite{taniguchi13APEX}. 
The averaged angle $\theta$ and the direct current injected into the STO are related by the following equation, 
\begin{equation}
\begin{split}
  I_{\rm dc}
  =&
  \frac{2\alpha e \lambda MV}{\hbar \eta \cos\theta}
  \left(
    \frac{1}{\sqrt{1-\lambda^{2}\sin^{2}\theta}}
    -
    1
  \right)^{-1}
\\
  &\times
  \left[
    H_{\rm appl}
    +
    \left(
      H_{\rm K}
      -
      4\pi M 
    \right)
    \cos\theta
  \right]
  \sin^{2}\theta.
  \label{eq:current}
\end{split}
\end{equation}
The physical meaning of Eq. (\ref{eq:current}) is that, when a direct current $I_{\rm dc}$ is injected, 
an auto-oscillation with the cone angle $\theta$ satisfying Eq. (\ref{eq:current}) is excited 
with the oscillation frequency of $f(\theta)$, where 
\begin{equation}
  f(\theta)
  =
  \frac{\gamma}{2\pi}
  \left[
    H_{\rm appl}
    +
    \left(
     H_{\rm K}
     -
     4\pi M
    \right)
    \cos\theta
  \right]. 
  \label{eq:frequency}
\end{equation}
Note that the averaged value of $\theta$ can be regarded as the tilted angle of the magnetization from the $z$ axis, 
whereas $\varphi$ is the phase of the magnetization in the $xy$ plane. 


On the other hand, in the presence of the alternating current, 
the spin torque due to the alternating current locks the frequency and phase of the STO 
when the condition 
\begin{equation}
  2\pi 
  \left[
    f(\theta)
    -
    f_{\rm ac}
  \right]
  =
  -\frac{\sqrt{\mathscr{A}^{2}+\mathscr{B}^{2}} \gamma H_{\rm ac}}{2 \sin\theta}
  \sin\left(\Phi-\phi^{\prime}\right), 
  \label{eq:locking_condition}
\end{equation}
is satisfied \cite{yamaguchi19}; see also Supplementary data. 
Here, we introduce the phase difference between the STO and the alternating current as 
\begin{equation}
  \Phi
  \equiv
  \varphi
  -
  2\pi f_{\rm ac} t, 
  \label{eq:phase_diff}
\end{equation}
where $2\pi f_{\rm ac}t$ is the phase of the alternating current, according to Eq. (\ref{eq:current_def}). 
Note that the phase difference $\Phi$ is constant in the synchronized state 
because the phase $\varphi$ oscillates with the frequency $f_{\rm ac}$ when the synchronization is realized. 
The dimensionless quantities $\mathscr{A}$ and $\mathscr{B}$ are given by 
\begin{equation}
  \mathscr{A}
  =
  \frac{\delta\omega(\theta) \sin\theta\cos\theta}{F(\theta)}
  \frac{2}{\lambda^{2}\sin^{2}\theta}
  \left(
    \frac{1}{\sqrt{1-\lambda^{2}\sin^{2}\theta}}
    -
    1
  \right),
\end{equation}
\begin{equation}
  \mathscr{B}
  =
  \frac{2 \left(1-\sqrt{1-\lambda^{2}\sin^{2}\theta}\right)}{\lambda^{2}\sin^{2}\theta}, 
\end{equation}
where 
\begin{equation}
  \delta 
  \omega(\theta)
  =
  \gamma
  \left(
    H_{\rm K}
    -
    4\pi M 
  \right)
  \sin\theta,
\end{equation}
\begin{equation}
\begin{split}
  F(\theta)
  =&
  \gamma 
  H_{\rm dc}
  \left[
    \frac{\lambda^{2}\cos^{2}\theta}{(1-\lambda^{2}\sin^{2}\theta)^{3/2}}
  \right.
\\
  &
  \left.
    -
    \frac{1}{\sin^{2}\theta}
    \left(
      \frac{1}{\sqrt{1-\lambda^{2}\sin^{2}\theta}}
      -
      1
    \right)
  \right]
\\
  &
  -
  \alpha
  \gamma
  \left[
    H_{\rm appl}
    \cos\theta
    +
    \left(
      H_{\rm K}
      -
      4\pi M
    \right)
    \cos 2\theta
  \right]. 
\end{split}
\end{equation}
The angle $\phi^{\prime}$ satisfies $\sin\phi^{\prime}=\mathscr{A}/\sqrt{\mathscr{A}^{2}+\mathscr{B}^{2}}$ and $\cos\phi^{\prime}=\mathscr{B}/\sqrt{\mathscr{A}^{2}+\mathscr{B}^{2}}$. 
Since $\lambda\sin\theta<1$, $\mathscr{A}$ and $\mathscr{B}$ are approximated as 
$\mathscr{A}\simeq \delta\omega(\theta)\sin\theta\cos\theta/F(\theta)$ and $\mathscr{B}\simeq 1$. 
Note that $|\mathscr{A}| \gg 1$ for typical parameters \cite{yamaguchi19}. 
Equation (\ref{eq:locking_condition}) indicates that the phase difference $\Phi$ is a function of the frequency of the alternating current $f_{\rm ac}$. 
Accordingly, measuring the diode voltage as a function of $f_{\rm ac}$ enables us to identify the phase difference $\Phi$, as shown below. 


Now let us investigate the role of the STO's phase on the spin-torque diode effect. 
The resistance of a magnetic tunnel junction is well described as 
\begin{equation}
  R
  =
  \frac{R_{0}}{2}
  -
  \frac{\Delta R}{2}
  \mathbf{m}
  \cdot
  \mathbf{p}, 
  \label{eq:resistance}
\end{equation}
where $R_{0}=R_{\rm P}+R_{\rm AP}$ and $\Delta R=R_{\rm AP}-R_{\rm P}$ 
with the resistances $R_{\rm P}$ and $R_{\rm AP}$ being the parallel and antiparallel alignments of the magnetizations. 
The second term of Eq. (\ref{eq:resistance}) shows the oscillation reflecting the magnetization oscillation in the free layer. 
In the injection-locked state, the oscillation frequency is identical to that of the alternating current. 
Therefore, the rectified voltage of the spin-torque diode effect is defined as \cite{tulapurkar05}
\begin{equation}
  V_{\rm dc}
  =
  \frac{1}{T}
  \int_{0}^{T}
  dt
  I_{\rm ac}
  \cos 
  \left( 2\pi f_{\rm ac} t \right) 
  \frac{-\Delta R}{2} 
  \mathbf{m}
  \cdot
  \mathbf{p}, 
  \label{eq:voltage_def}
\end{equation}
where $T=1/f_{\rm ac}$. 
Note that $\mathbf{m}\cdot\mathbf{p}=m_{x}=\sin\theta\cos\varphi$ in the present system. 
Since the tilted angle $\theta$ of the magnetization is almost constant in the auto-oscillation state, we find 
\begin{equation}
  V_{\rm dc}
  =
  -\frac{I_{\rm ac}\Delta R}{4}
  \sin\theta
  \cos\Phi. 
  \label{eq:diode_voltage}
\end{equation}
Substituting Eq. (\ref{eq:locking_condition}) to Eq. (\ref{eq:diode_voltage}) and using the fact that $|\mathscr{A}|\gg 1$ and $\mathscr{B} \simeq 1$, 
Eq. (\ref{eq:diode_voltage}) can be rewritten as 
\begin{equation}
  V_{\rm dc}
  \simeq
  I_{\rm ac}
  \Delta R
  \frac{\pi [f(\theta)-f_{\rm ac}] \sin^{2}\theta}{\gamma H_{\rm ac} \sqrt{1+\mathscr{A}^{2}}}.
  \label{eq:diode_voltage_1}
\end{equation}


Equations (\ref{eq:diode_voltage}) and (\ref{eq:diode_voltage_1}) predict several interesting features of the rectified voltage generated by an STO. 
For example, Eq. (\ref{eq:diode_voltage_1}) indicates that the dependence of the diode voltage on the frequency $f_{\rm ac}$ of the alternating current is linear; 
not a Lorentzian nor anti-Lorentzian function as in the case of the conventional spin-torque diode effect \cite{tulapurkar05}. 
The difference is due to the fact that the conventional spin-torque diode effect results from the FMR (linear oscillation) state, 
whereas the present study deals with a nonlinear oscillation. 
In addition, Eq. (\ref{eq:diode_voltage}) indicates that the diode voltage reflects the phase difference $\Phi$ between the STO and the alternating current. 
The result implies that the spin-torque diode effect of the STO 
can be used to estimate the oscillator's phase experimentally. 




\begin{figure}
\centerline{\includegraphics[width=1.0\columnwidth]{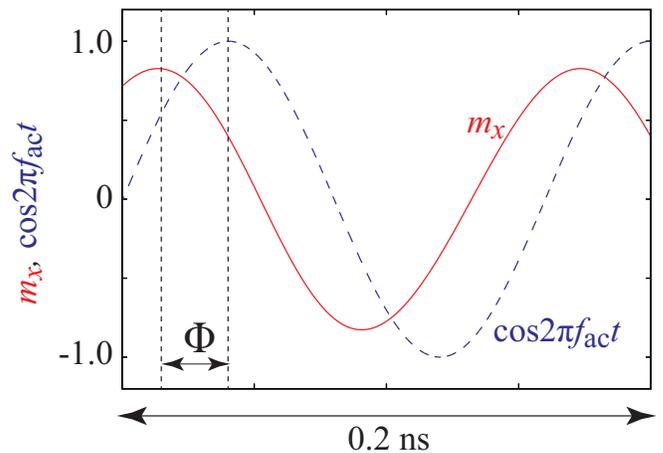}}
\caption{
         Time evolutions of $m_{x}$ and $\cos 2\pi f_{\rm ac}t$ for $I_{\rm ac}=0.03$ mA and $f_{\rm ac}=6.26$ GHz. 
         The phase difference $\Phi$ in this case is $60^{\circ}$. 
         \vspace{-3ex}}
\label{fig:fig2}
\end{figure}



We perform numerical simulation of Eq. (\ref{eq:LLG}) to investigate the validity of Eqs. (\ref{eq:diode_voltage}) and (\ref{eq:diode_voltage_1}). 
The values of the parameters used in the following are obtained from the experiment \cite{kubota13} and its theoretical analysis \cite{taniguchi13APEX} as 
$M=1448.3$ emu/c.c., $H_{\rm K}=1.8616 \times 10^{4}$ Oe, $H_{\rm appl}=2.0$ kOe, $V=\pi \times 60^{2} \times 2$ nm${}^{3}$, 
$\eta=0.537$, $\lambda=0.288$, $\gamma=1.764 \times 10^{7}$ rad/(Oe s), and $\alpha=0.005$. 
The resistance difference at the parallel and antiparallel alignment of the magnetizations is $\Delta R=150$ $\Omega$. 
The magnitudes of the direct and alternating currents are fixed to $I_{\rm dc}=2.5$ mA and $I_{\rm ac}=0.03$ mA, respectively. 
The oscillation frequency excited by this direct current is estimated to be $6.24$ GHz from the LLG simulation, 
corresponding to that the averaged tilted angle is about $\theta=56.9^{\circ}$. 


Figure \ref{fig:fig2} shows an example of the definition of the phase difference $\Phi$, 
where the time evolutions of $m_{x}$ and the alternating current [$\cos (2\pi f_{\rm ac}t)$] are shown. 
The frequency of the alternating current $f_{\rm ac}$ is set to be $f_{\rm ac}=6.26$ GHz.
The figure indicates that the frequency of the STO is fixed to that of the alternating current, and 
the phase difference in this case is nearly $60^{\circ}$. 



\begin{figure}
\centerline{\includegraphics[width=1.0\columnwidth]{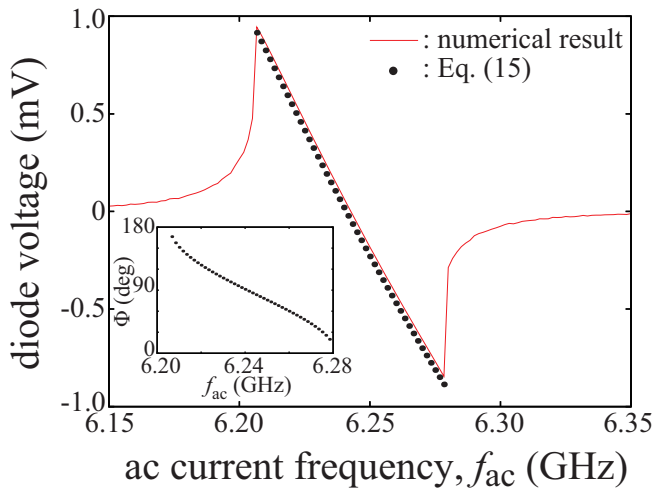}}
\caption{
         Dependence of the diode voltage on the frequency of the alternating current, where $I_{\rm ac}=0.03$ mA. 
         Solid line is obtained by evaluating Eq. (\ref{eq:voltage_def}) numerically with the numerical solution of $m_{x}$,
         whereas dots are obtained by using Eq. (\ref{eq:diode_voltage}) using the phase difference $\Phi$ estimated by the numerical simulation. 
         The inset shows the phase $\Phi$ as a function of $f_{\rm ac}$. 
         \vspace{-3ex}}
\label{fig:fig3}
\end{figure}




Next, we examine the validity of Eqs. (\ref{eq:diode_voltage}) and (\ref{eq:diode_voltage_1}) by the following approach. 
First, we evaluate the diode voltage defined by Eq. (\ref{eq:voltage_def}) with the numerical solution of $m_{x}$ obtained by the LLG simulation. 
The solid line in Fig. \ref{fig:fig3} shows the diode voltage obtained by this method. 
Note that a finite voltage appears when the injection locking is achieved. 
For the present parameters, the injection locking occurs for $6.21 \lesssim f_{\rm ac} \lesssim 6.28$ GHz. 
Outside the locking range, the diode voltage becomes nearly zero. 
This is because the oscillation frequency of the magnetization differs from that of the alternating current, 
and thus, a long-time average of Eq. (\ref{eq:voltage_def}) becomes zero, 
although the numerical simulation is performed during a finite time, and thus, the voltage in Fig. \ref{fig:fig3} remains finite. 
It should be emphasized that the diode voltage in the locking region shows a linear dependence on $f_{\rm ac}$, 
indicating the validity of Eq. (\ref{eq:diode_voltage_1}). 
Second, we compare this diode voltage with the theoretical formula given by Eq. (\ref{eq:diode_voltage}). 
The dots in Fig. \ref{fig:fig3} are obtained from Eq. (\ref{eq:diode_voltage}) by inserting the value of the phase difference estimated by the LLG simulation, as done in Fig. \ref{fig:fig2}. 
The inset of Fig. \ref{fig:fig3} shows the relation between the frequency of the alternating current and the phase $\Phi$ in the locked state.
The results indicate that the diode voltage in the frequency domain reflects the phase of the STO. 
In other words, the spin-torque diode effect of the STO is useful to estimate its phase. 


Although the numerical results are well explained by the analytical formulas, 
we need to validate the applicability of these formulas for completeness. 
An assumption used in the derivation of these formulas is that the cone angle $\theta$ of the magnetization oscillation is solely determined by the direct current through Eq. (\ref{eq:current}). 
Strictly speaking, however, the cone angle in the presence of the alternating current depends not only on $I_{\rm dc}$ but also on $I_{\rm ac}$ and $f_{\rm ac}$. 
The dependence of the diode voltage on $I_{\rm ac}$ and $f_{\rm ac}$ is rather complex. 
For example, Eq. (\ref{eq:diode_voltage}) with the assumption of $\theta$ being solely determined by direct current indicates that the dependence of the diode voltage on $f_{\rm ac}$ is linear. 
However, since $\theta$ in Eq. (\ref{eq:diode_voltage}) depends on $f_{\rm ac}$, the diode voltage is not a simple linear function of $f_{\rm ac}$. 
Simultaneously, however, we should emphasize that the real value of the cone angle is close to that estimated by Eq. (\ref{eq:current}), 
and therefore, our proposal to estimate the STO's phase from the spin-torque diode effect works well, as can be seen in Fig. \ref{fig:fig3}. 
The detail of these points is summarized in Supplementary data. 






It should also be noted that another direct voltage, $I_{\rm dc} R_{0}$, will appear in experiment \cite{zhang18}, in addition to Eq. (\ref{eq:diode_voltage}). 
However, this direct voltage can be experimentally separated from the rectified voltage 
because it is independent of the magnitude and frequency of the alternating current. 
Therefore, we consider that this contribution to the direct voltage does not affect the phase evaluation proposed in this work. 


In conclusion, the spin-torque diode effect of an STO was studied theoretically. 
An analytical formula of the diode voltage was derived, which indicates that the rectified voltage of the STO depends linearly on the frequency of the alternating current.  
The formula also reveals that the diode voltage depends nonlinearly on 
the phase difference between the magnetization and the alternating current injected into the STO. 
Numerical simulation of the LLG equation confirmed the validities of the analytical calculations. 
The result implied that measuring the spin-torque diode voltage of the STO is useful to evaluate the oscillator's phase. 


The authors are grateful to Yoshishige Suzuki and Minori Goto for valuable discussions. 
This paper was based on the results obtained from a project (Innovative AI Chips and Next-Generation Computing Technology Development/(2) 
Development of next-generation computing technologies/Exploration of Neuromorphic Dynamics towards Future Symbiotic Society) commissioned by NEDO. 



\end{document}